\documentclass[pra ,showpacs,showkeys,twocolumn]{revtex4-1}

\usepackage{graphicx}
\usepackage{latexsym}
\usepackage{amsmath}
\usepackage{amssymb}
\usepackage{amsfonts}
\usepackage{color}
\usepackage[normalem]{ulem}

\newcommand{\jj}{\mathbf{j}}
\newcommand{\xx}{\mathbf{x}}
\newcommand{\kk}{\mathbf{k}}
\newcommand{\Meff}{M_\textrm{eff}}

\begin{document}


\title{Classical field model for arrays of photon condensates}

\author{Vladimir N. Gladilin}
\author{Michiel Wouters}

\affiliation{TQC, Universiteit Antwerpen, Universiteitsplein 1,
B-2610 Antwerpen, Belgium}

\date{\today}

\begin{abstract}
We introduce a classical phasor model for the description of
multimode photon condensates that thermalize through repeated
absorptions and reemissions by dye molecules. Thermal equilibrium is
expressed through the fluctuation-dissipation relation that connects
the energy damping to spontaneous emission fluctuations. We apply
our model to a photonic Josephson junction (two coupled wells) and
to one- and two-dimensional arrays of photon condensates. In the
limit of zero pumping and cavity losses, we recover the thermal
equilibrium result, but in the weakly driven-dissipative case in the
canonical regime, we find suppressed density and phase fluctuations
with respect to the ideal Bose gas.
\end{abstract}

\maketitle

\section{Introduction}

Bose-Einstein condensates (BEC) of photons can be realized by
embedding dye molecules in a high-quality optical cavity
\cite{klaers10}. When the rate of absorption and reemission of
photons by the dye molecules is much higher than the photon losses,
the photons are brought to thermal equilibrium with the molecular
rovibrational states
\cite{kennard,stepanov,klaers_np10,moroshkin14}. These are in turn
thermalized by collisions with solvent molecules. The photon gas
then assumes the solvent temperature and features Bose-Einstein
condensation above the saturation density \cite{klaers10}.

Since the ideal Bose gas is one of the simplest systems from the
theoretical point of view \cite{huang}, one may wonder whether this
system presents any theoretical challenges or is just a nice
platform to perform some demonstration experiments of elementary
textbook physics. When the photon losses are fully absent, the
latter is the case \cite{kruchkov14,berman17}, since then the system
is guaranteed to relax to the thermal state. When on the other hand
losses cannot be fully neglected, as is the case in current
experiments \cite{klaers10,marelic16,greveling18,nyman18,dung17},
the physics becomes richer and our understanding of the interplay
between pumping, losses, external potentials and thermalization is
still not complete. Most experiments are performed with harmonically
trapped photons, but more recently there has been experimental
progress in the creation of double well and periodic potentials
\cite{dung17}.

The simplest theoretical description of  out of equilibrium 
photon condensation consists
of rate equations for the occupation of the single particle energy
levels \cite{klaers11}. The ingredient that is missing here is the
coherence between the single particle states, which is
necessary to form localized photon wave packets \cite{kirton16}.
Such a state is for example formed when the photon condensate is
pumped with a finite size pumping spot \cite{schmitt15}. An
extension of the rate equation model to take the spatial
distribution of the molecules into account was developed by Hesten
et al. \cite{hesten18}.

On the other size of the complexity spectrum are the quantum optics
based approaches where the master equation for the open quantum
system \cite{breuer} consisting of the coupled molecular and
photonic system is solved \cite{kirton13,kirton15,kirton16}. Apart
from the needed computational resources, a disadvantage of this
master equation approach is that it is hard to include correlation
between photons and molecules. It has been shown that the
correlation between photons and molecules can affect the density
fluctuations \cite{klaers12,schmitt14}. The underlying physics is
elementary: when a large number of molecules is present, they form a
bath for particle exchange as in the grandcanonical statistical
ensemble. In this so-called grandcanonical regime, the number
fluctuations are large, due to the unrestricted exchange of
particles between system and bath. When on the other hand the number
of molecules is small, it becomes unlikely that all photons are
simultaneously absorbed; hence number fluctuations are reduced. In
order to describe this physics correctly, it is essential to include
the correlations between the number of photons and the number of
excited molecules. From the quantum optics side, these classical
molecule-photon correlations can be included most easily in a
quantum trajectory approach. For single mode photon condensates,
this was implemented in Ref. \cite{verstraelen19}. It was also shown
in this work that the quantum optical description is well reproduced
by a classical field model.

For the description of close to ideal Bose gases, classical field
theory has proved to be an indispensable tool \cite{proukakis17}.
Most experiments with weakly interacting ultracold Bose gases are
excellently described by the Gross-Pitaevskii equation (GPE)
\cite{bec_book}. For exciton-polariton condensates, a system closely
related to photon condensates, many experiments are modeled with a
generalized Gross-Pitaevskii equation that includes pumping and
losses \cite{carusotto13}. Exciton-polaritons are hybrid
light-matter quasiparticles that interact with each other thanks to
their excitonic component, setting them apart from noninteracting
photons. In practice however, the interaction energy in
experimentally realized polariton condensates is quite small due the
relatively small value of the interaction constant \cite{delteil19}.
In parallel to photon condensates, cavity losses make it necessary
to pump the system in order to reach a steady state. When cavity
losses are small, thermal equilibrium is closely approached
\cite{sun17}.

Classical field theories exist at various levels of complexity. The
most elementary version is the standard GPE, containing only kinetic
and interaction energy, that is suitable for the description of zero
temperature weakly interacting bosons \cite{bec_book}. Particle
exchange with a reservoir can be added by including an imaginary
term \cite{carusotto13}. Energy exchange between the bosons and
their environment can be modeled by making the prefactor of the time
derivative complex. It was originally introduced to model the
friction between the superfluid and normal components of liquid
Helium \cite{pitaevskii59}, but has also been employed in the
description of ultracold atoms \cite{proukakis17,konabe06} and
polariton condensates \cite{wouters12}. The standard
Gross-Pitaevskii equation assumes perfect coherence of the bosons.
Decoherence can be incorporated by including some stochasticity.
This can for example be derived in the truncated Wigner
approximation \cite{quantumnoise,sinatra02,proukakis17,wouters09}.

For photon condensates, a classical field description has been used
to model their phase coherence \cite{schmitt16}. The `phasor model'
version of the classical field description was developed in the
context of laser physics \cite{henry82,scully97} and compares
favorably with a quantum trajectory description for single-mode
photon condensates \cite{verstraelen19}. It is the purpose of this
paper to extend the phasor model to multimode photon condensates.

We show in Sec. \ref{sec:model} that the thermalization of the
photon gas by the molecules is described by adding the same term
that models the friction between superfluid and normal components in
atomic condensates. The fluctuations in the phasor model are shown
to be related to this friction through a fluctuation-dissipation
relation.
In Sec. \ref{sec:1mode}, we recapitulate the physics of single mode
photon condensates, with specific attention to the regimes of small
and large density fluctuations, the `canonical' and `grandcanonical'
regimes respectively. We then analyze in detail the case of two
coupled photon traps in Sec. \ref{sec:pjj}, a photonic Josephson
junction (PJJ). For this system, we analytically compute the density
and phase fluctuations in the linearized Bogoliubov approximation.
In the limit of zero losses, we recover the equilibrium correlators
in the classical regime, justifying our model as an adequate
description of photon condensates. Finally, in Sec.
\ref{sec:lattices}, we  extend our analysis to 1D and 2D lattices of
photon condensates. We find that the spatial coherence of photon
condensates is much better in the canonical regime as compared to
the grandcanonical regime.

\section{Model \label{sec:model}}

\subsection{Kennard-Stepanov relation and energy relaxation}

\begin{figure} \centering
\includegraphics[width=0.8\linewidth]{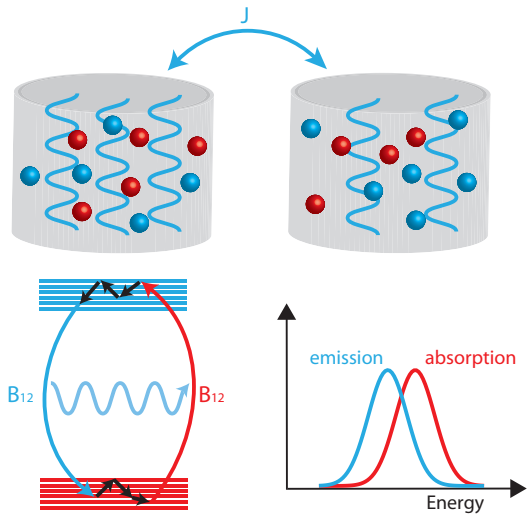}
\caption{ We consider an array of cavities coupled by photon
tunneling. Photons are emitted and reabsorbed by the dye molecules.
The molecules undergo scattering with the solvent molecules, which
thermalizes the occupations of the rovibrational states. The
emission and absorption coefficients satisfy the Kennard-Stepanov
relation.  } \label{fig:sketch}
\end{figure}

For dye molecules that interact sufficiently strongly with their
solvent, the emission and absorption coefficients are related by the
Kennard-Stepanov (KS) law, which reads at the inverse temperature
$\beta=1/(k_B T)$ \cite{kennard,stepanov,moroshkin14}
\begin{equation}
\frac{B_{12}}{B_{21}} = e^{\beta (\omega-\omega_0)}. \label{eq:KS}
\end{equation}
Here $B_{12}$ ($B_{21}$) is the Einstein coefficient for absorption
(emission) of a photon, $\omega_0$ is the molecular transition
frequency and $\omega$ is the photon frequency. 
Here we absorbed possible degeneracy factors in a  renormalization of the 
molecular transition frequency. A
sketch of an absorption-emission spectrum that satisfies the KS
relation is shown in Fig. \ref{fig:sketch}.

The KS law brings the photon gas to thermal equilibrium, as can be
seen from the steady state of the kinetic equation for the photon
number in a single mode cavity:
\begin{equation}
\frac{d n}{dt} = -B_{12} n + B_{21}(n+1).
\end{equation}
Setting $dn/dt=0$, one finds $n=(B_{12}/B_{21}-1)^{-1}$, which
reduces to the Bose-Einstein distribution when the KS relation
\eqref{eq:KS} is used.

For a single mode of noninteracting photons, the KS relation can be
straightforwardly implemented in a theoretical model, but for
multimode systems, the photonic frequency $\omega$ is not \textit{a
priori} known. When the photonic frequency is still close to a
certain cavity frequency $\omega_c$, one can proceed by writing the
KS relation as
\begin{equation}
\frac{B_{12}(\omega)}{B_{21}(\omega)}  =e^{\beta \Delta} e^{\beta
(\omega-\omega_c)},
\end{equation}
where $\Delta = \omega_c-\omega_0$ is the cavity-molecule detuning.
To be specific, we will assume that all the energy dependence is in
the absorption coefficient. This is valid close to the maximum of
the emission. However, we do not expect that our results will be
significantly altered when the energy dependence is moved to the
emission or distributed between absorption and emission. Choosing
$\omega_c$ as the zero of energy and replacing $\omega \rightarrow i
\partial_t$, one obtains
\begin{align}
B_{12} = B_{21} e^{\beta \Delta} (1+i\frac{\partial}{\partial t}).
\label{eq:B12_omega}
\end{align}

In a classical field description, the photon dynamics is described
by a generalized Gross-Pitaevskii equation (gGPE), setting
$\hbar=1$,
\begin{equation}
i \frac{\partial \psi}{\partial t}  = \hat T \psi +
\frac{i}{2}(B_{21}M_2-B_{12} M_1 - \gamma) \psi. \label{eq:gGP}
\end{equation}
Here, $\hat T$ formally represents the kinetic energy and $M_{1(2)}$
are the number of ground state (excited) molecules. The cavity loss
rate is denoted by $\gamma$. The wave function is position dependent
(explicitly: $\psi=\psi(\jj)$, where $\jj$ labels a lattice cite),
as well as the number of ground state and excited molecules, that
satisfy at all times $M_1(\jj)+M_2(\jj)=M$, where $M$ is the number
of dye molecules at each lattice site. Eq. \eqref{eq:gGP} does not
contain interaction energy, which is quite negligible in current
experiments \cite{randonji18}, except for a slow thermo-optical
nonlinearity \cite{alaeian17}, inclusion of which would be a
straightforward extension of our model.

With Eq. \eqref{eq:B12_omega}, the gGPE becomes
\begin{equation}
i(1+i\kappa) \frac{\partial \psi}{\partial t}  = \hat T \psi +
\frac{i}{2}B_{21}(M_2 - e^{\beta \Delta} M_1-\gamma)\psi ,
\end{equation}
where
\begin{equation}
\kappa= \frac{1}{2} \beta B_{21}e^{\beta \Delta} M_1 \label{kappa}
\end{equation}
is the energy relaxation rate. For typical photon condensates the
relative fluctuations in the number of ground state molecules is
small, such that $\kappa$ can be approximated by a constant.

The above derivation made use of the formal substitution $\omega
\rightarrow i \frac{\partial}{\partial t}$ in the KS relation, whose
validity may be questioned. In order to further justify this
approach, we show in Appendix \ref{app:twomode} that the same
equation can be rigorously derived for the case where the energy
dependent absorption is due to coupling with a lossy bosonic mode.

For $\kappa\ll 1$, which is satisfied if the absorption of photons
is slow on the thermal time scale $\beta$, the gGPE can be
approximated by
\begin{equation}
i \frac{\partial \psi}{\partial t}  = (1-i\kappa) \hat T \psi +
\frac{i}{2}B_{21}(M_2 - e^{\beta \Delta} M_1-\gamma)\psi ,
\label{eq:gGPE}
\end{equation}
where $i\kappa \hat T \psi$ forms an imaginary tunneling term, while
corrections due to $\kappa$ in the emission-absorption term were
neglected.

\subsection{Fluctuations}
A classical field model without fluctuations only captures
absorption and stimulated emission. In order to describe spontaneous
emissions, fluctuations have to be introduced. One possibility is
the heuristic phasor model, where spontaneous emissions are modeled
by adding a unit length phasor with random angle to the photonic
field \cite{henry82}:
\begin{equation}
\psi(\jj) \rightarrow \psi(\jj) + e^{i\theta}, \label{eq:dpsi_spont}
\end{equation}
with $\theta$ a random phase. This noise should be added at random
times, with probability $p=dt\; B_{21} M_{\uparrow 2}(\jj)$ in a
time interval $dt$. For the single mode case, this model was
demonstrated to show excellent agreement with a full quantum
trajectory description~\cite{verstraelen19}, motivating its use to
describe fluctuations in the multimode regime.

\subsection{Molecule dynamics}

The equation of motion \eqref{eq:gGPE} has to be coupled to the
dynamics of the number of excited molecules. The absorption and
stimulated emission dynamics are local and lead to a local change in
the number of excited molecules of the form
\begin{equation}
\left. \frac{\partial M_2(\jj)}{\partial t}\right|_{\rm
abs\,+\,st.\,em.} = B_{21}(e^{\beta \Delta} M_1(\jj) -
M_2(\jj))|\psi|^2. \label{eq:abs_stim}
\end{equation}
From the energy relaxation term (proportional to $\kappa$), we have
the contribution
\begin{equation}
\left.\frac{\partial M_2(\jj)}{\partial t}  \right|_{\rm relax} = 2
\kappa\, \textrm{Re}[\psi^*(\jj) (\hat T \psi)(\jj)],
\label{eq:relax}
\end{equation}
where we have neglected relaxation of the molecules in other modes
than the cavity mode. Finally, the spontaneous emission
\eqref{eq:dpsi_spont} is accompanied by a change
\begin{equation}
\left. d M_{2}(\jj)\right|_\textrm {sp.\,em.} =
-(|\psi(\jj)+e^{i\theta}|^2-|\psi(\jj)|^2). \label{eq:spont}
\end{equation}
Note that this change can be positive or negative and is in a given
realization not equal to one (this is only true on average). The
terminology `spontaneous emission' may therefore be a bit confusing
for this term. The crucial physics that it does capture are the
phase diffusion  and density fluctuations in the photon condensate
\cite{verstraelen19}.

In order to compensate for the excitations, which are lost through
the cavity mirrors, and reach a steady state, the system has to be
continuously pumped. This is modeled by the following term in the
equation of motion for the excited molecules
\begin{equation}
\left. d M_{2}(\jj)\right|_\textrm {pump} = \gamma \bar n +
\sqrt{\gamma \bar n} \xi_p,
\end{equation}
where $\bar n$ is the targeted steady state number of photons. The
Gaussian white term with autocorrelation $\langle \xi_p(t) \xi_p(t')
\rangle = \delta(t-t')$ comes from  the shot noise in the excitation
of the molecules. We include it here for completeness, but it will
turn out that its effect is much smaller than that of the
spontaneous emission noise \eqref{eq:spont}.

\section{Single mode physics: canonical and grandcanonical regimes
\label{sec:1mode} }

The noninteracting Bose gas in the grandcanonical ensemble has large
number fluctuations. When the photons of a photon condensate are
coupled to a large number of molecules, the molecules form a
reservoir and photon number fluctuations are large. With less
molecules in the cavity, the photon number fluctuations are reduced.

The dynamical analysis of the density fluctuations for a single-mode
photon condensate as a function of reservoir size and detuning was
performed in Ref. \cite{verstraelen19}. The deviations of the number
of photons $n$ and total number of excitations $X=M_2+n$ from their
equilibrium values $\bar n$ and $\bar X$ evolve in linear
approximation as
\begin{align}
\frac{d }{dt} \delta X &=-\gamma\, \delta n   + \sqrt{\gamma \bar n} \xi_p, \label{linsm1} \\
\frac{d }{dt} \delta n &= -\Gamma\, \delta n +
\frac{\sigma_n^2}{\Meff} \Gamma\, \delta X + \sqrt{2 B_{12} M_1 \bar
n} \xi_n.
 \label{eq:linsm2}
\end{align}
Here, the number fluctuation decay rate is given by \cite{schmitt18}
\begin{equation}
\Gamma = \left(1+\frac{n^2}{\Meff}\right)\frac{B_{21} M_2}{\bar n},
\label{eq:Gamma}
\end{equation}
where the effective reservoir size equals
\begin{equation}
\Meff = \frac{ M+\gamma e^{-\beta \Delta}/B_{21}}{2[1+\cosh(\beta
\Delta)]}
\end{equation}
This reduces to the form from Ref. \cite{schmitt14} when $\gamma=0$.

The Gaussian white noises $\xi_{p,n}$ have zero mean and variance
equal to $\langle \xi_i(t) \xi_j(t') \rangle =\delta(t-t')
\delta_{i,j}$. The stochastic term $\xi_n$ originates from the
fluctuations due to spontaneous emission and can be derived in the
diffusion approximation to the phasor model by considering the
effect of the spontaneous emission on the variance of the density.
In a spontaneous emission $\psi = \psi + e^{i\theta}$, the density
variance increases by
\begin{equation}
\Delta \textrm{Var}[n]= \langle (2 |\psi| \cos \theta)^2 \rangle = 2
n.
\end{equation}
With the spontaneous emission rate being $B_{21} M_2$, one arrives
at the noise term in Eq. \eqref{eq:linsm2}. Analogously, the noise
term in Eq. \eqref{linsm1} originates from the shot noise deviations
of the pumping, which is needed to compensate for the photon losses,
from its average rate $\gamma \bar n$.

According to Eq. \eqref{eq:linsm2},  the density fluctuations in the
absence of losses are given by \cite{verstraelen19}
\begin{equation}
\sigma_n^2 = \frac{\Meff \bar n^2}{\Meff + \bar n^2}, \label{eq:dn2}
\end{equation}
Typical experimental photon losses do not significantly alter the
density fluctuations \cite{schmitt14}. In the limit of a large
reservoir (grandcanonical regime) ($\bar n^2 \ll M_\textrm{eff}$),
one obtains large number fluctuations $\sigma_n^2= \bar n^2$, where
in the opposite limit of a small reservoir (canonical regime), one
obtains small number fluctuations $\sigma_n^2 = M_\textrm{eff} \ll
\bar n^2$. In the grandcanonical limit, phase jumps occur when the
density goes to zero \cite{schmitt16}, but the overall phase
coherence time is still of the Schawlow-Townes form
\cite{scully97,deleeuw14}.

\section{The photonic Josephson junction \label{sec:pjj}}

The simplest system to illustrate our model for a lattice of photon
condensates is a photonic Josephson juntion (PJJ), which consists of
two coupled sites (L and R) \cite{dung17}. For this example, we will
write explicitly
 the gGPE and relaxation contribution to the molecular dynamics for the left site; the equations for the right
site can be obtained by the replacement $L \leftrightarrow R$.

\subsection{Equations of motion}
The deterministic part of the equations of motion for the field
amplitude on the left hand site reads from Eq. \eqref{eq:gGP}
explicitly
\begin{equation}
i \frac{\partial}{\partial t} \psi_L = - J(1-i\kappa)\psi_R +
\frac{i}{2}  B_{21}\left( M_{2 L} - e^{\beta \Delta} M_{1 L} -\gamma
\right) \psi_L. \label{eq:jos1}
\end{equation}
The relaxation coefficient $\kappa$ leads to a larger linewidth for
the antisymmetric state as compared to the symmetric state. It leads
to the following change in the number of excited molecules, cf. Eq.
\eqref{eq:relax},
\begin{equation}
\left.\frac{d M_{2 L,R}}{dt}\right|_{\rm relax}= - 2 \kappa J
\textrm{Re} \left(\psi_L^* \psi_R\right).
\end{equation}
For the remaining equations of motion, we refer to Eqs.
\eqref{eq:dpsi_spont}, \eqref{eq:abs_stim} and \eqref{eq:spont}.

\subsection{Fluctuations}

As in the single mode case, further analytical insight can be
obtained by linearizing the equations of motion for small phase and
density difference, writing $\psi_j=\sqrt{\bar n+\delta n_j}e^{i
\theta_j}$. For the phase difference $\Delta \theta =
\theta_R-\theta_L$, one then obtains
\begin{align}
\frac{\partial}{\partial t} \Delta \theta &= - 2 \kappa J \Delta
\theta -\frac{J}{\bar n} \Delta n + \sqrt{2D_{\Delta \theta}}
\xi_\theta, \label{eq:phase_dyn}
\end{align}
where the phase diffusion originates from spontaneous emission. The
phase diffusion constant is
\begin{align}
D_{\Delta \theta} =   \frac{M_2 B_{21}}{2\bar n}. \label{eq:Ddtheta}
\end{align}
Eq.~\eqref{eq:phase_dyn} shows that the energy relaxation parameter
$\kappa$ drives the system to zero relative phase. With
Eq.~\eqref{kappa}, the phase damping and noise are seen to obey the
fluctuation-dissipation relation
\begin{equation}
D_{\Delta \theta} = k_B T \frac{\kappa}{\bar n}. \label{eq:fdrel}
\end{equation}
For the photon and excitation number density difference $\Delta n =
\delta n_R- \delta n_L$ and $\Delta X = \delta X_R- \delta X_L$, one
obtains
\begin{align}
\frac{d }{dt} \Delta X &=4 J \bar n \Delta \theta-\gamma\, \Delta n   + \sqrt{2 \gamma \bar n} \xi_p, \label{lin1} \\
\frac{d }{dt} \Delta n &= 4 J \bar n \Delta \theta -(\Gamma + 2 \kappa J )\, \Delta n \nonumber \\
 &+ \frac{\sigma_n^2}{\Meff} \Gamma\, \Delta X
 + \sqrt{4 B_{12} M_1 \bar n} \xi_n,
\end{align}
where $\Gamma$ and $\sigma_n^2$ are still given by Eqs.
\eqref{eq:Gamma} and \eqref{eq:dn2}.

With Ito calculus, equations of motion for the correlation functions
can be constructed and the correlators can be evaluated
analytically. The full expressions are cumbersome, but both in the
limit for large and for small tunneling, the variance of the phase
difference takes the simple expression
\begin{equation}
\langle (\Delta \theta)^2 \rangle = \eta \frac{D_{\Delta \theta} }{4
\bar n \kappa J}, \label{eq:dtheta1}
\end{equation}
where $\eta \geq 1$ is a noise enhancement factor that describes the
increase of fluctuations due to losses:
\begin{equation}
\eta= 1+\frac{\gamma e^{-\beta \Delta}}{B_{21} M}. \label{eq:eta}
\end{equation}
In practice, this factor becomes appreciably larger than one only
for large negative detuning.

Inserting in Eq. \eqref{eq:dtheta1} the fluctuation-dissipation
relation \eqref{eq:fdrel}, one finds
\begin{equation}
\langle (\Delta \theta)^2 \rangle = \eta \frac{k_B T}{2 \bar n  J}.
\label{eq:phasebog}
\end{equation}
For the relative density fluctuations, one obtains in the limit of
large $J$
\begin{equation}
\frac{\langle (\Delta n)^2 \rangle}{\bar n^2}= \eta \frac{2 k_B
T}{\bar nJ} \label{eq:g2eqbog}
\end{equation}
In the absence of losses ($\gamma=0$), the results
\eqref{eq:phasebog} and \eqref{eq:g2eqbog} agree with those for
noninteracting bosons in the weak fluctuation regime at thermal
equilibrium (see appendix \ref{app:pjj_gc}) . Note that this
limiting equilibrium expression is ensured by the
fluctuation-dissipation relation, which originates from the
Kennard-Stepanov relation.

From the density and phase fluctuations, also the first order
coherence can be computed in linearized approximation:
\begin{equation}
\frac{\langle \psi_L^\dag \psi_R \rangle}{\bar n} = 1-\frac{\langle
(\Delta \theta)^2 \rangle}{2}- \frac{\langle (\Delta n)^2 \rangle}{8
\bar n^2}. \label{eq:coh_2w}
\end{equation}
In the limit of large $J$ one obtains
\begin{equation}
\frac{\langle \psi_L^\dag \psi_R \rangle}{\bar n} = 1-\eta \frac{k_B
T}{2 \bar n  J}
\end{equation}
which also reduces to the equilibrium expression when $\eta=1$.

\begin{figure} \centering
\includegraphics[width=0.9\linewidth]{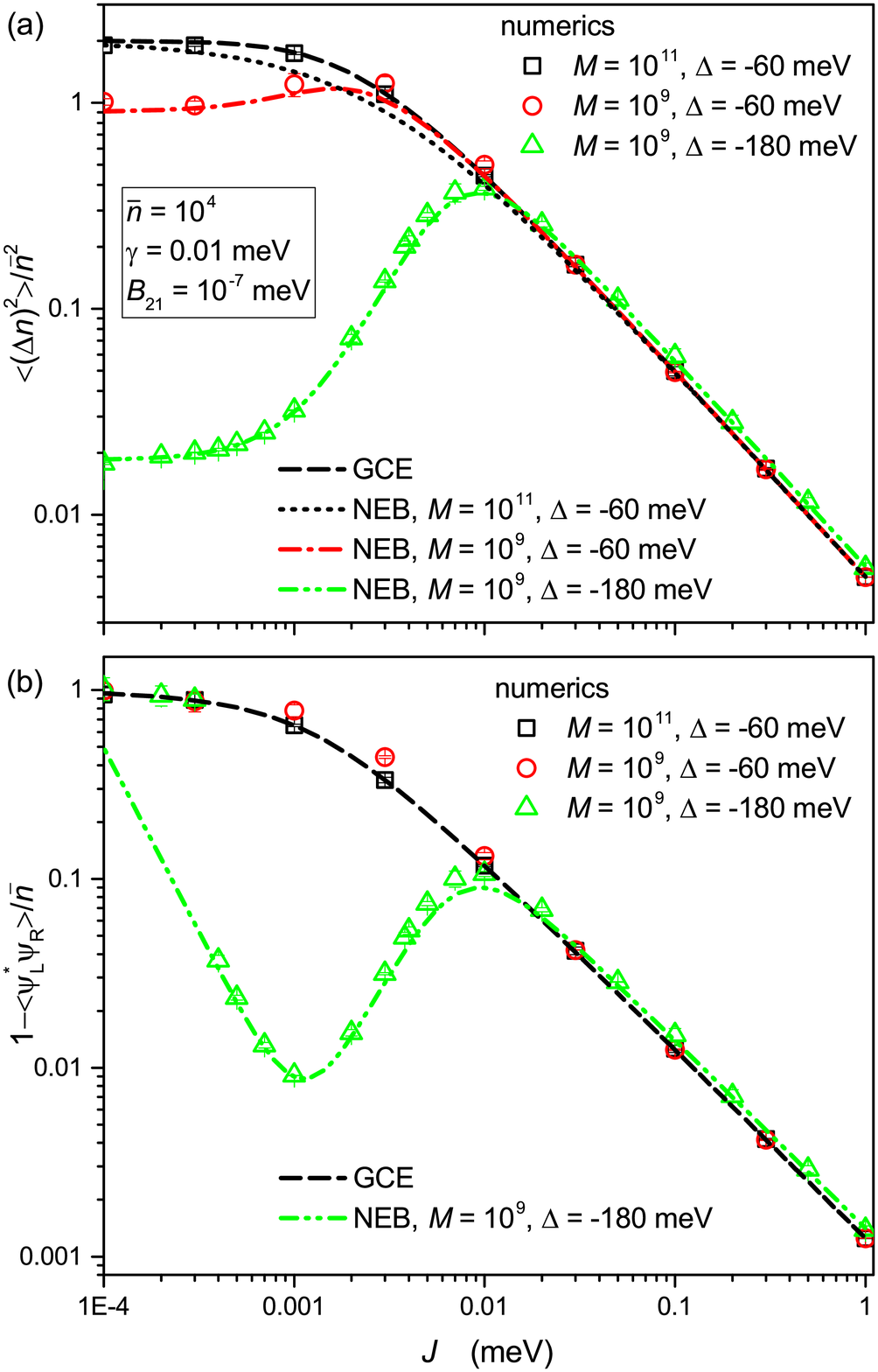}
\caption{(a) Relative density fluctuations and (b) complement to the
first order coherence in a photonic Josephson junction as a function
of the tunneling amplitude $J$ for several values of the effective
reservoir size (black dashed and dotted lines and squares: deep
grandcanonical, red dash-dotted line and circles: intermediate,
green dash-dot-dot line and triangles: deep canonical). The symbols
were obtained with a direct numerical simulation of the phasor
model. The red dotted, green dash-dot-dot and black dotted lines
were obtained with the linearized Bogoliubov theory. The black
dashed line was obtained with the equilibrium theory of
noninteracting bosons in the grandcanonical ensemble.
\label{fig:2cavF1n}}
\end{figure}


The dependence of density fluctuations on the tunneling rate in the
presence of losses is shown in Fig. \ref{fig:2cavF1n}(a) for photon
condensates both in the grandcanonical (black squares), canonical
(green triangles) and intermediate (red dots) regimes. These points
were obtained from direct numerical simulations of the equations of
motion.

At sufficiently large tunneling rate, the density fluctuations in
Bogoliubov approximation \eqref{eq:g2eqbog} are recovered for all
photon and molecule numbers. The Bogoliubov approximation breaks
down when density fluctuations are too large (at small $J$), but in
equilibrium, the density fluctuations can be computed analytically
at any $T/J$ (see appendix \ref{app:pjj_gc}). This expression is
shown in Fig. \ref{fig:2cavF1n}(a) with the black dashed line and
corresponds very well to the numerical simulations in the deep
grandcanonical regime for all tunneling rates.

In the canonical regime, losses affect the density fluctuations
significantly. This is seen both in the numerics (red and green
symbols) and in the nonequilibrium Bogoliubov (NEB) expression (red
and green lines). The density fluctuations in the canonical regime
can be understood from the interplay between tunneling and losses.
By coupling the two wells, the relative density can fluctuate
because of particle exchange. When the tunneling becomes large, the
condensate is almost entirely in the symmetric state, such that the
relative density fluctuations become small. On the other hand, when
the effect of tunneling is much smaller than that of losses,
particle exchange is barely possible and relative density
fluctuations are suppressed. Consequently, there is a nonmonotoneous
dependence of the relative density fluctuations on the tunneling
rate. By combining the small and large $J$ expansions of the density
fluctuations, the position of the maximal density fluctuations can
be estimated to be at the tunneling rate $J=\frac{1}{2}(T \gamma
B_{21} n^2/M_2)^{1/3}$.

The first order coherence of the PJJ is shown in Fig.
\ref{fig:2cavF1n}(b). At large tunneling, it shows the same behavior
as the density fluctuations. Also in analogy with density
fluctuations, the coherence for grandcanonical condensates is almost
unaffected by experimentally relevant losses. In the canonical
regime, on the other hand, we find a again a nonmonotonous behavior.
At the tunneling strength where density fluctuations become
suppressed, also the first order coherence improves. Upon further
decreasing the tunneling strength, the coherence becomes optimal and
then worsens again. This worsening at small $J$ is entirely due to
phase fluctuations. Indeed, in the limit of zero tunneling, the
phases between the two condensates become uncorrelated, such that
there is no coherence between the two condensates. From the small
$J$ expansion of the coherence, the maximum is found to be at the
tunneling rate $J=[B_{21} n \gamma/(4 \sqrt{2})]^{1/2}$.

In some numerical simulations (not shown here), we have also
observed a long lived antisymmetric state with $\pi$ phase
difference between the two wells. This is not the lowest energy
state, but for small tunneling rate the energy relaxation is too
weak in order to cause a dynamical instability of the antisymmetric
state. The metastability of the antibonding state is relevant for
the potential use of photonic Bose-Einstein condensates as analog
simulators for (classical) minimization problems
\cite{isingmachine}: already for the simple two-site problem, there
appears the possibility for the system to be stuck in an excited
state. A more detailed analysis of metastable states of photon
condensates is beyond the scope of this paper and will be deferred
to another study.

\begin{figure} \centering
\includegraphics[width=0.9\linewidth]{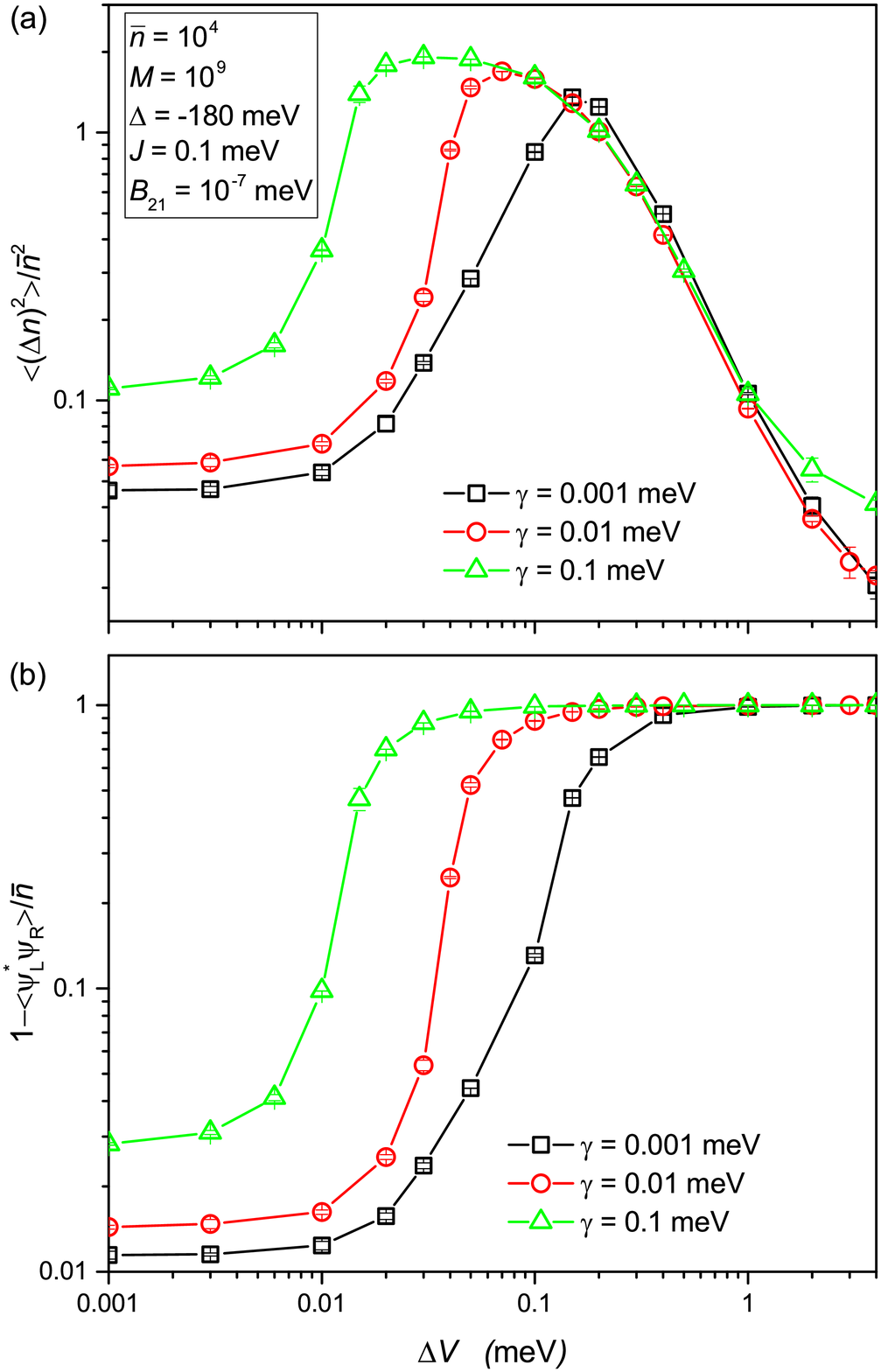}
\caption{(a) Density fluctuations as a function of the energy offset
between the two wells in the canonical regime. (b) The behavior of
the coherence between the two wells as a function of the energy
offset between the two wells in the canonical regime.}
\label{fig:2cavF3nTEMP}
\end{figure}

The previous discussion assumed perfect symmetry of the system. In
Fig. \ref{fig:2cavF3nTEMP}, we show the sensitivity of the density
fluctuations and coherence to the detuning of the wells in the
canonical regime. Where the spatial coherence is very good for small
energy offset between the wells, it quickly deteriorates when the
detuning is increased [see Fig. \ref{fig:2cavF3nTEMP}(b)]. For
larger losses, the coherence is lost for smaller values of the
detuning. The reason is that the system goes to a desynchronized
state, where the condensates in the two wells have different
frequencies and hence no phase coherence. When losses are larger,
less particles can be transferred between the two wells and
synchronization is lost for smaller detuning. Also the density
fluctuations are sensitive to a detuning between the wells [see Fig.
\ref{fig:2cavF3nTEMP}(a)]. When the system loses synchronisation,
the density fluctuations increase, but for larger detuning, the
density fluctuations again decrease, the exchange of photons being
suppressed when the detuning is much larger than the hopping.

\section{Lattices of photon condensates \label{sec:lattices}}

With the physics of the two-site PJJ understood, we now turn to the
study of one- and two-dimensional arrays. From the analytical side,
we will again study the dynamics in Bogoliubov approximation, which
allows to compute the spatial coherence. We complete these
calculations with numerical studies of one- and two-dimensional
systems.

The linear analysis for larger lattices can be performed along the
same lines as for the PJJ. The density and phase variables are again
introduced as $\psi(\jj) = \sqrt{\bar n + \delta n(\jj)} e^{i
\delta\theta(\jj)}$. The Fourier components of the phase
fluctuations are defined by
\begin{equation}
\delta \theta (\jj) = \frac{1}{\sqrt{L}} \sum_\kk \delta
\theta_\kk\; e^{i \kk \cdot \jj},
\end{equation}
where $L$ is the number of lattice sites, and analogous for $\delta
n(\jj)$ and $\delta X(\jj)$.

As observables, we will consider the momentum distribution
\begin{equation}
N_{\kk} = \langle \psi^\dag_{\kk} \psi_{\kk} \rangle,
\end{equation}
where $\psi_{\kk}$ is the Fourier transform of the field
$\psi(\jj)$. It is worth stressing that the momentum distribution
$N_{\kk}$ is different from the Fourier transform of the density
$\delta n_\kk$. We will also consider the normalized static
structure factor
\begin{align}
S_{\kk} &= \frac{1}{\bar n^2} \sum_\jj \langle \delta n(\jj) \delta n(0) \rangle e^{-i \jj \cdot \kk}  \\
&= \frac{1}{\bar n^2}  \langle |\delta n_\kk|^2 \rangle
 \label{eq:Sk}
\end{align}

In the linear Bogoliubov approximation to Eqs.
(\ref{eq:gGPE})-(\ref{eq:spont}), the fluctuations obey the
equations of motion
\begin{align}
\frac{\partial}{\partial t} \delta \theta_\kk &= -\kappa \epsilon_\kk \delta \theta_\kk -\frac{\epsilon_\kk}{2n} \delta n_\kk + \sqrt{2 D_\theta } \xi^{(\theta)}_\kk \label{eq:dthetak} \\
\frac{\partial}{\partial t} \delta n_\kk &= 2 \bar n \epsilon_\kk
\delta \theta_\kk -(\Gamma + \kappa \epsilon_\kk )\delta n_\kk +
\sqrt{2 D_n } \xi^{(n)}_\kk \label{eq:dnk}
\\
\frac{\partial}{\partial t} \delta X_\kk &= 2 \bar n \epsilon_\kk
\delta \theta_\kk  - \gamma \delta n_\kk + \sqrt{2 D_X}
\xi^{(p)}_\kk. \label{eq:dXk}
\end{align}
For a tight binding hamiltonian with hopping amplitude $J$, the
single particle dispersion equals $\epsilon_\kk =
2J[2-\cos(k_x)- \cos(k_y)]$. The white noise terms have zero average
and variance $\langle \xi^{(\alpha)}_\kk(t) \xi^{(\beta)}_{\kk'}(t')
\rangle = \delta_{\kk,-\kk'} \delta_{\alpha,\beta} \delta(t-t')$.
The diffusion constants are $D_\theta = B_{21} M_2/4 \bar n$,
$D_n=B_{21} M_2 \bar n$ and $D_X=\gamma \bar n$. As in the
two-cavity case, the last one appears to play a negligible role.
These equations are identical to the ones in the case of the PJJ
with the replacement $2J \rightarrow \epsilon_\kk$. The momentum
distribution can be computed analogously to Eq. \eqref{eq:coh_2w} as
\begin{equation}
N(k) = \frac{ \bar n}{2} \langle |\delta \theta_\kk|^2 \rangle +
\frac{1 }{8\bar n}\langle |\delta n_k|^2 \rangle.
\end{equation}

\begin{figure} \centering
\includegraphics[width=0.9\linewidth]{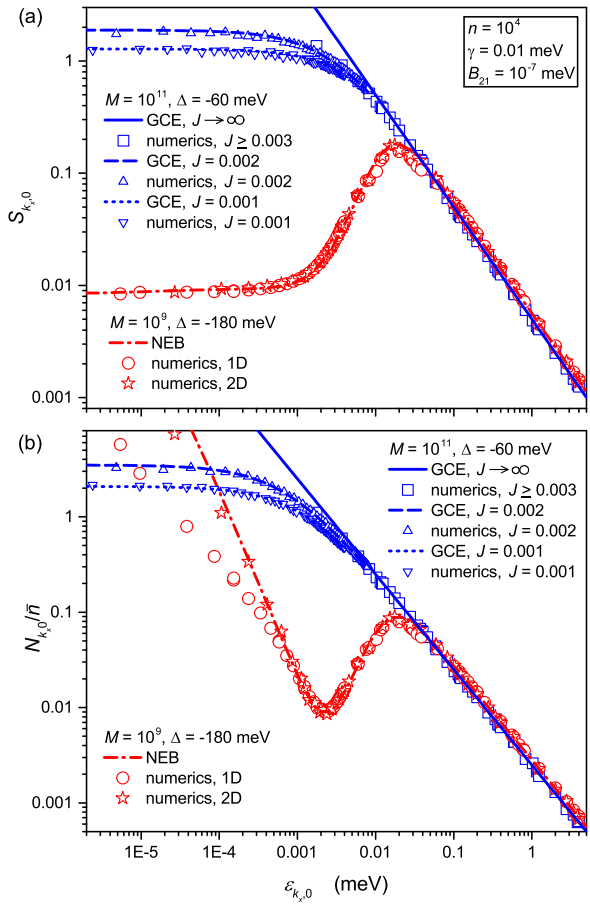}
\caption{(a) Static structure factor and (b) momentum distribution
of a photon condensate in a lattice as a function of the energy.
Symbols and lines were obtained from numerics and analytical
calculations respectively. Blue squares and triangles were obtained
in the 1D grandcanonical regime. Red circles and stars refer to
simulations in the canonical regime in 1D and 2D, respectively. }
\label{fig:arrF1nk}
\end{figure}

Numerically, we can check the validity of the linear approximation
through the static structure factor, shown in Fig.
\ref{fig:arrF1nk}(a) and the momentum distribution, shown in Fig.
\ref{fig:arrF1nk}(b). In order to stress the similarity with the
PJJ, we plot the fluctuations $S_{k_x,0}$ and $N_{k_x,0}$ against
the dispersion $\epsilon_\kk$. The parameter values used in the
figures correspond to a 1D grandcanonical (blue squares and
triangles), a 1D canonical (red circles) and 2D canonical (red
stars) regimes.

In order to span the five orders in magnitude in energy, simulations
on systems with different tunneling were combined. As can be seen
from Eqs. \eqref{eq:dthetak}-\eqref{eq:dXk}, in the linear regime,
the fluctuations only depend on the energy $\epsilon_\kk$ and not on
the specific value of $J$. We verified that in the overlapping
energy regions, numerical simulations with different $J$ gave the
same results. This universality breaks down in the regime of strong
fluctuations, where the momentum distribution and static structure
factor explicitly depend on $J$ at small energy (compare upright and
inverted blue triangles).

In the grandcanonical regime at low energy, the fluctuations become
large and the linear approximation breaks down. In analogy with the
PJJ, one can use the grandcanonical equilibrium ensemble in order to
describe the condensate fluctuations in the grandcanonical regime
(see Appendix \ref{app:lattice_gc}). The corresponding blue dashed
and dotted lines closely reproduce the numerical results at all
energies.

In the large energy limit, we obtain from the nonequilibrium
Bogoliubov approximation the limiting expressions
\begin{align}
N(\kk) &= \eta \frac{k_B T} { \epsilon_\kk}          &(k\rightarrow \infty), \\
S(\kk) &= 2 \eta \frac{k_B T} {\bar n \epsilon_\kk}  &(k\rightarrow
\infty).
\end{align}

\begin{figure} \centering
\includegraphics[width=0.9\linewidth]{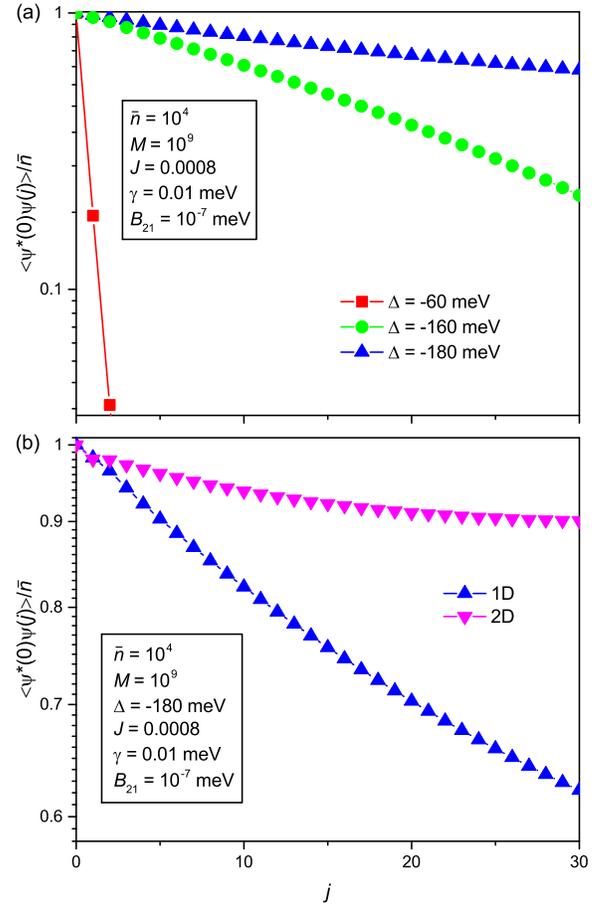}
\caption{(a) Real space spatial coherence in a 1D lattice for
various values of the detuning. (b) Comparison between real space
spatial coherence in a 1D (blue upward triangles) and 2D (pink
downward triangles) array with system parameters in the canonical
regime. The size of the simulation region is $L=128$ for 1D and
$L_x=L_y=64$ for 2D.} \label{fig:arrF2}
\end{figure}

In order to gain insight in the behavior of correlations at very
large distances, one can start from the low energy dependence in
Bogoliubov approximation of the phase-phase correlator:
\begin{equation}
\langle |\delta \theta_\kk|^2 \rangle = \eta \frac{k_B T}{2 \bar  n
\epsilon_\kk}  \hspace{1cm} (k\rightarrow 0), \label{eq:phase_k}
\end{equation}
The divergence of the phase fluctuations at small energy is
reflected in the momentum distribution (Fig. \ref{fig:arrF1nk}(b)).
It appears that at the lowest energies, there is a discrepancy
between the results of the numerical simulations and the NEB
prediction. Moreover, the numerical results appear to depend on the
dimensionality (see the discrepancy between the 1D and 2D results),
where the linear theory is dimension independent. The 2D simulations
appear to follow the analytical curve somewhat longer, but in both
cases, the fluctuations in the numerical simulations are smaller
than the analytical prediction.

When density fluctuations are small, the phase correlator
\eqref{eq:phase_k} results in an exponential decay of the first
order spatial coherence at large distances, $\langle \psi^\dag(x)
\psi(x') \rangle \sim \exp(-|x-x'|/\ell_c)$, with coherence length
$\ell_c = 4 \bar n J /k_B T$. This expression coincides with the
correlation length of the 1D \textit{interacting} Bose gas
\cite{bec_book}. In equilibrium systems, the condition of small
density fluctuations is satisfied thanks to interactions, in the
present case of noninteracting bosons out of equilibrium, the
density fluctuations can be suppressed by losses. This was
illustrated in Fig. \ref{fig:arrF1nk} (red circles, stars and
dash-dotted line). If the main contribution to the momentum
distribution comes from the large momentum tail (for not too large
systems and not too small $J$), the decay of the first order spatial
correlation function is exponential. For a 1D lattice in the limit
where $\bar n J \gg k_B T$, the coherence length (in units of the
lattice spacing) equals $\ell_c=2\bar nJ/(\eta k_BT)$. In the
equilibrium limit $\eta=1$, it reduces to the correlation length of
the ideal Bose gas \cite{petrov04} (see Appendix \ref{app:lattice_gc}). From the
above analytical considerations, one expects an increase by a factor
of two in the correlation length when going from the grandcanonical
to the canonical regime.

The spatial coherence in real space, obtained from numerical
simulations in a 1D array, is shown in Fig. \ref{fig:arrF2}(a). From
these results obtained for various values of the detuning $\Delta$,
it is seen that the spatial coherence improves when going from the
grandcanonical {(red squares)}  to the canonical {(blue triangles)}
regime. It is also clear from this figure that the increase in the
spatial coherence length between grandcanonical (red squares) and
canonical (blue triangles) regimes is much larger than the
analytically predicted factor of two. The larger than expected
coherence reflects the fact that the numerical momentum distribution
in the canonical regime in Fig. \ref{fig:arrF1nk}(b) lies below the
analytical prediction. The theoretical understanding of the
coherence length in the canonical regime requires further
investigation.

Moreover, in our numerical simulations with small tunneling rates,
we have observed that, in analogy with the two-well case, the system
can spend a long time in states with large phase difference between
neighboring wells. We also defer a study of the occurrence of states
with large phase difference to a further study.

In one dimension the suppression of density fluctuations only
quantitatively affects the spatial coherence, where in two
dimensions the difference is qualitative. Where the ideal 2D bose
gas (which has large density fluctuations) does not feature a phase
transition to a phase coherent state at finite temperature (the
decay of coherence is exponential at all nonzero temperatures), the
interacting Bose gas (with suppressed density fluctuations) features
a Berezinskii-Kosterlitz-Thouless transition.

In the nonequilibrium case, we have shown that density fluctuations
can remain small for noninteracting photons in the canonical regime
when losses are present (see Fig. \ref{fig:arrF1nk}(a)). The phase
correlator \eqref{eq:phase_k} then leads to an algebraic decay of
the spatial coherence \cite{bec_book}
\begin{equation}
\langle \psi^\dag(\xx) \psi(\xx') \rangle \propto |\xx-\xx'|^{-\nu}
\label{eq:coh_x2D}
\end{equation}
with the exponent
\begin{equation}
\nu = \eta \frac{k_B T}{4\pi \bar n J}.
\end{equation}
This prediction for the long distance spatial coherence reduces to
the equilibrium one when $\eta \rightarrow 1$.

Fig. \ref{fig:arrF2}(b) shows a comparison of the spatial coherence
between a one and a two dimensional  array. As predicted by our
theoretical analysis, the spatial coherence is better in the
two-dimensional case.

\section{Conclusions and outlook}

We have introduced a classical model to describe Bose-Einstein
condensates of photons in coupled cavities. The model consists of a
photon field that is coupled to the molecular states by emission and
absorption. The transfer of photons between the cavities is
described by the usual nearest neighbor tunneling term. The
interplay between tunneling and thermalization through the repeated
absorption and emission processes is modeled with an imaginary
tunneling term. This term was derived in the approximation that the
temperature is much larger than the tunneling energy. If this is not
the case, a higher order expansions of the Kennard-Stepanov relation
has to be made.

We have studied numerically the full nonlinear equations of motion
and obtained analytical expressions from the linearization of the
model around a homogeneous steady state, both for two coupled wells
and for lattices. From our analytical solutions, we have recovered
the equilibrium expressions in the limit of zero cavity losses.

In the grandcanonical ensemble with large density fluctuations, our
numerical simulations coincide with the thermal equilibrium results,
even when losses are present. In the canonical regime with small
density fluctuations, on the other hand, the cavities decouple from
each other when the tunneling rate is reduced. This results in a
nonmonotoneous dependence of the relative density fluctuations on
the tunneling rate.

For one- and two-dimensional lattices of photon condensates in the
grandcanonical regime, the spatial coherence reduces to that of the
ideal bose gas at thermal equilibrium. In the canonical regime, the
spatial coherence markedly improves. In the one-dimensional case,
the decay is still exponential, but with a longer coherence length.
Where the analytical Bogoliubov analysis predicts an enhancement of
the coherence length by a factor of two, the numerical simulations
show a much larger coherence length. Further theoretical work will
be needed to understand the spatial coherence in the canonical
regime. In 2D lattices in the canonical regime, the coherence decays
much slower than in 1D. From our Bogoliubov analysis, we predict a
power law decay, but the numerics is not conclusive on this point
because of the limited lattice sizes that were accessible in our
simulations. If the power law coherence exists, this would open the
way to the observation of the Berezinksii-Kosterlitz-Thouless
transition in (noninteracting) photon condensates, that is
stabilized by driving and dissipation
{\cite{caputo2016,wachtel,sieberer,gladilin19}.}

For the PJJ, we have shown that a detuning between the wells is
detrimental for the spatial coherence. It will be interesting to
analyze the role of disorder in extended lattices as well and study
the interplay between Anderson localization and driving and
dissipation
A further outlook concerns the study of photon BECs in lattices with
complex tunneling phases, where the engineering of artificical gauge
fields {\cite{tomoki}} could be possible.

\section{Acknowledgements}

Stimulating discussions with M. Weitz, J. Klaers, F. \"Ozt\"urk and
W. Verstraelen are warmly acknowledged. This work was financially
supported by the grant UA-BOF-FFB150168.

\appendix

\section{Energy relaxation from coupled bosonic modes \label{app:twomode}}

In order to motivate further the introduction of an energy
relaxation term for a system with energy-dependent losses, we derive
it also for the case of two coupled bosonic modes, where one
($\psi$) is conserved and one ($\chi$) is dissipative. Due to its
Lorentzian spectrum, the losses in the dissipative mode introduces
losses in the conserved one that are energy dependent. At the
classical field level, this system is described by
\begin{align}
i \frac{\partial}{\partial t} \psi &= \hat T \psi + g \chi, \label{eq:2m1} \\
i \frac{\partial}{\partial t} \chi &= \epsilon \chi + g \psi -
\frac{i}{2} \Gamma \chi.
\end{align}
For a constant $\hat T$ (no kinetic energy) and strong detuning
($|\epsilon-\hat T| \gg g,\Gamma$) it has an eigenvalue
\begin{equation}
\omega \approx \hat T - \frac{g^2}{(\hat T-\epsilon)^2}
\frac{i\Gamma}{2},
\end{equation}
showing an energy ($\hat T$)-dependent absorption rate. Expanding
around $\hat T=0$, one has
\begin{equation}
\omega \approx \hat T - \frac{g^2}{\epsilon^2} \frac{i\Gamma}{2} + i
\hat T \frac{g^2 \Gamma}{\epsilon^3 },
\end{equation}
from which one sees that the damping rate has the energy-dependent
form $\gamma=\gamma_0 + 2\kappa \hat T$ with
\begin{equation}
\gamma_0=  \frac{g^2 \Gamma}{\epsilon^2} \;\; \textrm{and} \;\;
\kappa = -\frac{g^2 \Gamma}{\epsilon^3}. \label{app:kappa}
\end{equation}

Let us now show that this system can be well approximated by a gGPE
with energy relaxation parameter $\kappa$. The equation of motion
for $\chi$ can be formally solved as
\begin{equation}
\chi=\frac{g \psi}{i \partial_t -\epsilon + i \Gamma/2}.
\end{equation}
Substituting in Eq. \eqref{eq:2m1}, one obtains
\begin{equation}
i \frac{\partial}{\partial t} \psi = \hat T \psi +  \frac{g^2 }{i
\partial_t -\epsilon + i \Gamma/2} \psi,
\end{equation}
After expansion to first order of the denominator in $\partial_t$
and subsequently in $\Gamma$, the equation for $\psi$ becomes
\begin{align}
i \frac{\partial}{\partial t} \psi &= \hat T \psi
-\frac{i}{2}\frac{g^2 }{\epsilon^2} \Gamma \psi
+ i \frac{g^2 \Gamma }{\epsilon^3} i \partial_t \psi, \\
& =  \hat T \psi - i \gamma_0 \psi - i \kappa \partial_t \psi,
\end{align}
where in the last line the definitions \eqref{app:kappa} of
$\gamma_0$ and $\kappa$ were used. We recover here the same relation
between the energy dependence of the loss rate and the gGPE as in
our derivation based on the KS relation.

\section{Grandcanonical treatment of the PJJ \label{app:pjj_gc}}
For two coupled wells, the total number of photons reads in terms of
the chemical potential and in the limit $k_B T \gg J$
\begin{equation}
2 \bar n=\frac{T}{-\mu}+ \frac{T}{2J-\mu}
\end{equation}
($\bar n$ is the number of photons in one cavity).

For the first order coherence, one finds
\begin{equation}
\langle \psi^\dag_L \psi_R \rangle  = \frac{1}{2} \left(
\frac{T}{-\mu}-\frac{T}{2J-\mu} \right).
\end{equation}
The relative density fluctuations can be computed by using Wick's
theorem, yielding
\begin{equation}
\langle (\Delta n)^2 \rangle = \frac{2 T^2}{(2J-\mu)(-\mu)}.
\end{equation}
In the limit where $\bar nJ\gg k_B T$, the first order coherence
reduces to $\langle \psi^\dag_L \psi_R \rangle  =\bar{n} -
\frac{T}{2J}$ and the density fluctuations reduce to $\langle
(\Delta n)^2 \rangle/\bar n^2= 2 k_B T/(\bar{n} J) $.

\section{Grandcanonical treatment of a 1D lattice \label{app:lattice_gc}}

The chemical potential of a noninteracting condensate in a lattice
can be determined from
\begin{equation}
\bar n = \int_0^{2\pi}  \frac{d k}{2\pi} \frac{k_B
T}{2J(1-\cos(k))-\mu} = \frac{k_B T}{\sqrt{\mu(\mu-4J)}},
\end{equation}
where the classical approximation to the Bose-Einstein distribution
was made, which is valid when $k_B T \gg J$. The momentum
distribution is given by the Bose-Einstein distribution
\begin{equation}
N_k = \frac{k_B T}{\epsilon(k)-\mu}.
\end{equation}
From the Fourier transform of the momentum distribution, one obtains
the spatial coherence, which decays exponentially at large
distances, with coherence length $\ell_c=2nJ/(k_BT)$. The static
structure factor can be computed by substituting
\begin{equation}
\delta n_k =  \sum_q \psi^\dag(q) \psi(k+q)
\end{equation}
in Eq. \eqref{eq:Sk} and then using Wick's theorem to obtain
\begin{equation}
S_k = \int_0^{2\pi} \frac{dk}{2\pi} \frac{(k_B T)^2}{\bar
n^2[\epsilon(q)-\mu][\epsilon(q+k)-\mu]},
\end{equation}
which was numerically evaluated to obtain the blue curves in Fig.
\ref{fig:arrF1nk}.

\end{document}